\documentclass{aa}

\usepackage[dvips]{graphicx}
\usepackage{verbatim, subfigure}
\usepackage{apalike} 
\usepackage{amsmath} 
\usepackage{dcolumn}
\usepackage{longtable}

\newcommand{\feh}{$[{\rm Fe/H}]$}

\newcommand{\dss}{$\delta$~Scuti stars}            
\newcommand{\ds}{$\delta$~Scuti star}
\newcommand{\ngc}[1]{NGC~{#1}}                              

\newcommand{\templogg}{{\sc templogg}}            
\newcommand{\vald}{{\sc vald}} 
\newcommand{\atlasni}{{\sc atlas9}} 
\newcommand{\atlasseks}{{\sc atlas6}}

\newcommand{\teff}{\ensuremath{T_{\rm eff}}}             
\newcommand{\logg}{\ensuremath{\log g}}                     
\newcommand{\vsini}{\ensuremath{v\sin i}}                   
\newcommand{\halpha}{H$\alpha$}  
\newcommand{\hbeta}{H$\beta$}
\newcommand{\halfa}{H$\alpha$}
 \newcommand{\kms}{\ensuremath{\text{km\;s}^{-1}}}           

\newcolumntype{p}{D{+}{\pm}{-1}}       

\newcommand{\task}{\texttt}
\newcommand{\ew}{EW}
 \newcommand{\vwa}{{\sc vwa}}           

\newlength{\dplotwidth}
\makeatletter
  \newcounter{liniet@l}[table]
  { \newcommand{\br}%
  {\stepcounter{liniet@l}\ifthenelse{\value{liniet@l}=#1}{\setcounter{liniet@l}{0}\\ \hline}{\\ }}%
   \begin{footnotesize}\begin{tabular}{#2}\hline \hline}%
  {\hline \hline \end{tabular}\end{footnotesize} }
\makeatother

\begin{document}
   \title{Rotation of stars in NGC 6134\thanks{Based in part on
           observations obtained at the European Southern Observatory at
           La Silla and UTSO-Las Campanas, Chile}}

   \subtitle{A comparison of \dss\ and non-variable stars}

   \author{M.~B.~Rasmussen
          \inst{1}
          \and
          H.Bruntt\inst{1}
          \and
          S.~Frandsen\inst{1}
	\and        
	E.~Paunzen\inst{2}
	\and
	 H.~M.~Maitzen\inst{2}}

   \offprints{S. Frandsen, \email{srf@ifa.au.dk}}

   \institute{Institute for Physics and Astronomy, 
               University of Aarhus, Bygn.~520, 
                DK-8000 Aarhus C, Denmark
	\and
	{Institut f\"ur Astronomie der Universit\"at Wien,
	T\"urkenschanzstrasse 17, A-1180 Wien, Austria}    }

\date{Received 14 February 2002 / Accepted 24 April 2002}
\abstract{We present results of spectroscopic observations of selected stars 
in the southern open cluster NGC~6134. We have determined the
rotational velocities of the six known \dss\ in 
NGC~6134 as well as several other non-variable stars 
with similar colour temperature in order to investigate
if \vsini\ and variability is somehow connected: we 
find no such correlation. We also compare the distribution 
of \vsini\ of \dss\ and  non-variable stars 
with four other well-studied open clusters to look for any 
systematic behaviour, but we find no conclusive evidence 
for \vsini\ and variability to be connected.
We have also used the spectra to carry out an abundance 
analysis of the \dss\ in NGC~6134 to confirm the 
high metal content of the cluster. We find 
$[{\rm Fe/H}] = +0.38\pm0.05$ which is in agreement with the
result obtained from Str\"omgren photometry.
We also present $\Delta$a photometry of the cluster, but we find no 
chemical peculiar stars based on this index.
 }

\maketitle

%

\section{Introduction\label{sec_int}}

The open cluster NGC 6134 has six confirmed \ds\ members (Frandsen et al. 1996) of
which five are multiperiodic. High quality Str\"omgren photometry 
has been carried out by Bruntt et al.\ (1999), except that the $c_1$ index 
(requiring $u$ images) could not be obtained with the CCD detector.
This paper describes a continuation of the earlier programs. We 
use spectroscopy and $\Delta$a-photometry 
(Maitzen 1976, Maitzen \& Vogt 1983) to obtain more
parameters and additional information about the stars in the cluster. 
These parameters are needed for the modeling and interpretation
of the oscillations in the $\delta$ Scuti stars. Here we will derive
rotational velocities (\vsini) for variable and non-variable stars and
carry out abundance analysis for the \dss.

\begin{figure*}\centering
  \begin{center}
    \subfigure[The non-stellar background signal is very high 
   in this case\label{subfig:blowup}]
{\includegraphics[width=\dplotwidth,
        bb=110 256 476 508, clip]{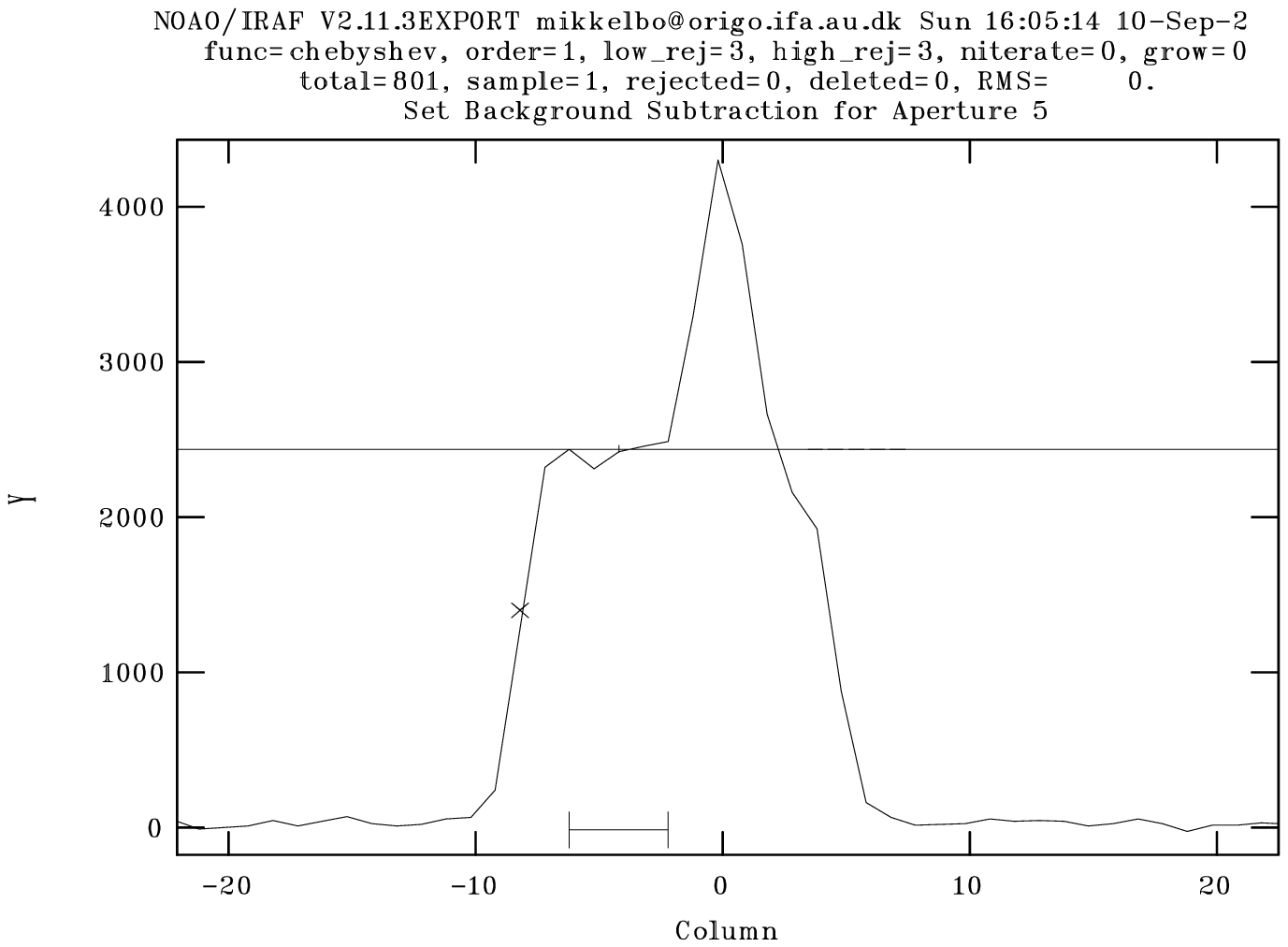}}
\qquad 
    \subfigure[For this star the background level is still significant
    but not dominating]{\includegraphics[width=\dplotwidth,
        bb=110 256 476 508, clip]{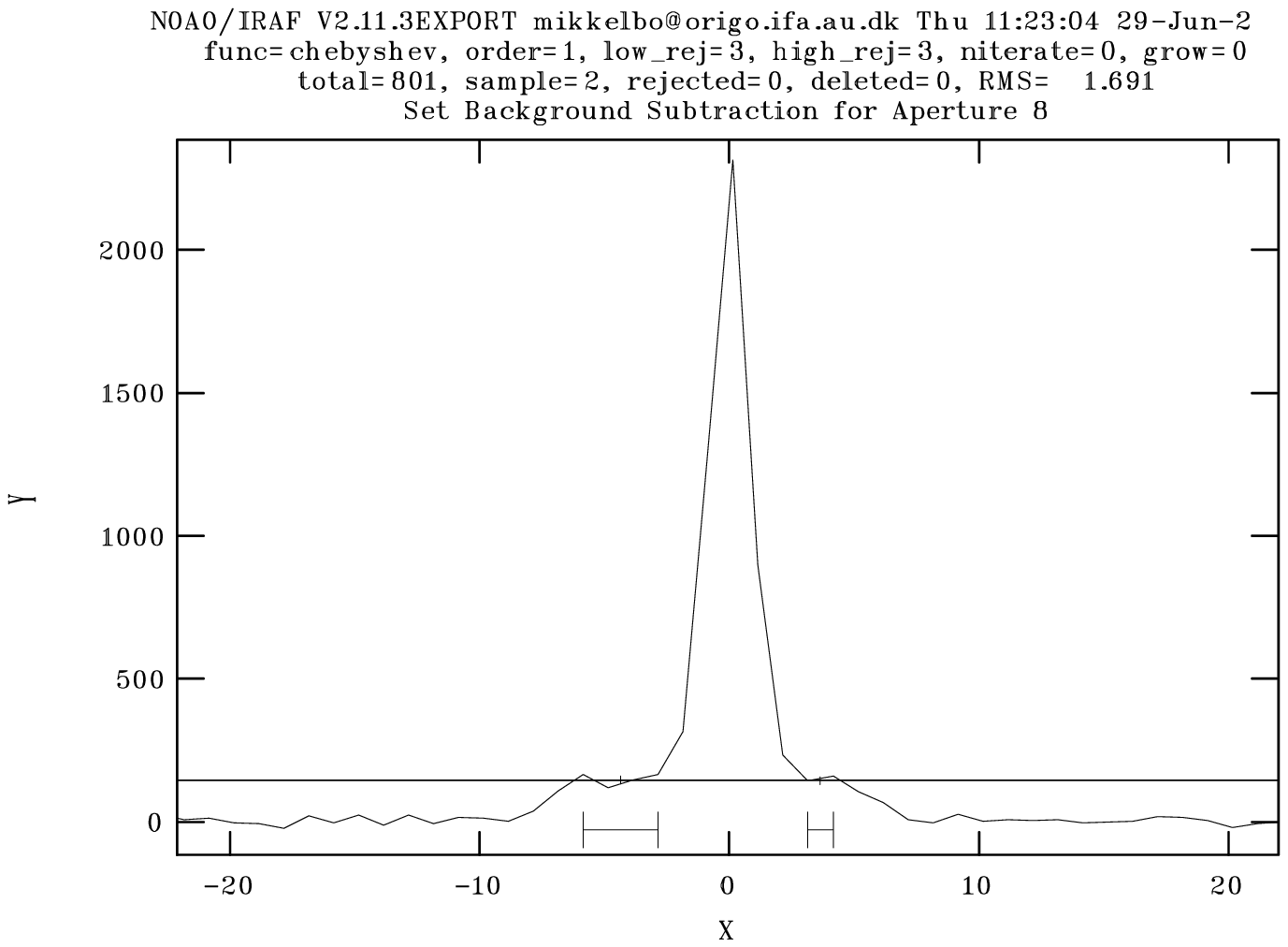}}
    \caption{Expanded plot of the cross section of an order for two
    different stars.
    The $x$-axis measures the distance
    from the center of the order in pixels. The horizontal bars in the
    lower part of each plot show how we have set the background
    regions in IRAF during manual editing of the apertures when using
    the task \task{apall}. In \ref{subfig:blowup} the high background signal
    indicates that the Moon is nearby\label{backgr}}
  \end{center}
\end{figure*}

In principle, with the combination of the information from the
frequencies of the \dss\ and the constraints imposed on
the variables by being members of an open cluster 
we will be able to make a very powerful test of stellar evolution.
The analysis of \dss\ is very complicated in stars that are fast rotators, 
so it is important
to locate slow rotators before extensive campaigns are started to
obtain oscillation spectra with very low noise levels, leading to
the detection of many oscillation frequencies. For example, the open cluster
Praesepe has 14 \dss\ and all but one rotate fast ($v \sin i > 100$~\kms) 
(Rodr\'{\i}guez et al.\ 2000).         
This has complicated the
determination of the stellar parameters considerably and made the
identification of the observed frequencies with radial or non-radial
modes very difficult (P\'erez Hern\'andez et al.\ 1999, 
Kjeldsen et al.\ 1998). 
The situation is not quite as bad for the Pleiades and  
Hyades clusters where a few slowly rotating \dss\ are found.

The nature of the excitation mechanism in \dss\ is an 
unsolved problem -- especially the understanding of 
which among the many possible modes predicted from models have
obserable amplitudes.
We have looked for systematic differences between
variable and non-variable stars. When the variables were discovered,
a large set of non-variable stars (amplitude $< 1$~mmag) were identified.
It is therefore possible to find pairs of stars, variable and non-variable
(twins), with similar photometric parameters. 
The \dss\ are located in the instability strip
and intrinsically unstable due to the opacity mechanism; but the
\dss\ have modes that do not seem to follow any regular pattern.
Many stars in the instability strip do not vary at all.
Perhaps the explanation is the difference in observables 
like \feh\ and \vsini\ among \dss\ and 
non-variable stars in the instability strip. We will examine this hypothesis
by comparing variable and non-variable stars with approximately the
same colour-temperature and evolutionary state.

\begin{figure*}\centering
  \includegraphics[width=\textwidth]{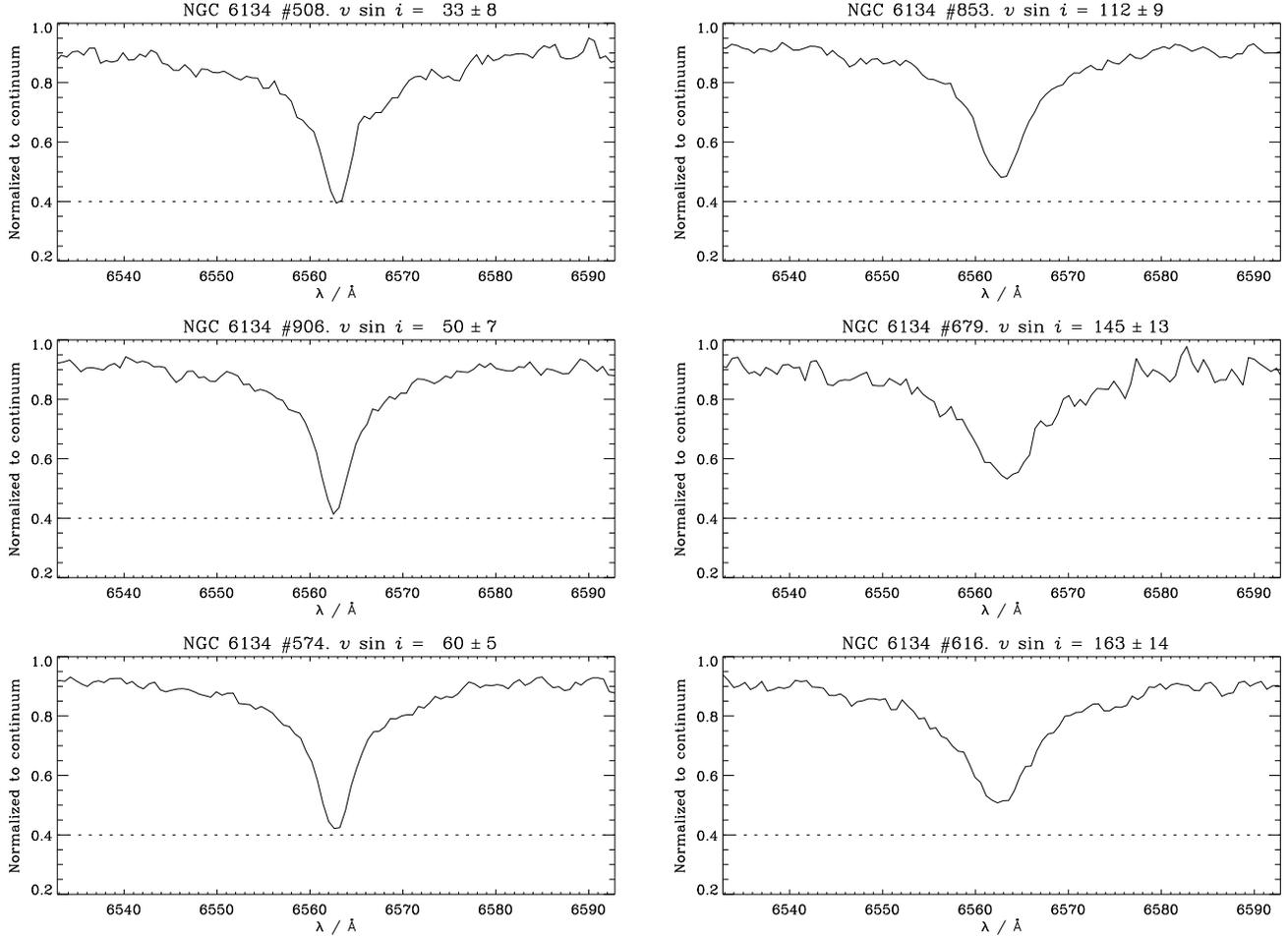}
  \caption[plot_ha.pro]{\halpha\ of the six \dss\ plotted for comparison.
  The dashed line at 0.4 is just a visual aid\label{spectra}}
\end{figure*}

Finally, we try to confirm the metallicity determined from 
Str\"omgren photometry (Bruntt et al.\ 1999)
and to locate possible Am or Ap-type stars among 
the $\delta$ Scuti stars. We will do abundance analysis of the 
\dss\ and we also investigate any possible chemical 
peculiarity with $\Delta$a photometry. 
It seems that the diffusion leading to chemically 
peculiar stars is coupled with slow rotation and seems to
exclude variability (Kurtz 2000).
But since NGC~6134 is a metal rich cluster, with a large number 
of A-type stars with high values of the 
Str\"omgren $m_1$ index -- also in case
of the \dss\ -- it would be interesting if some of 
these are CP-type stars.

The observations are described in 
Section~\ref{sec_obs} and in 
Section~\ref{sec_red} we describe the reduction of the spectra.
The calibration of the stellar parameters is given in 
Section~\ref{sec_cal} and in
Section~\ref{sec_rot} we determine the rotational velocities.
The abundance analysis is carried out in Section~\ref{sec_abu}.
The data reduction and results of the $\Delta$a photometry is given in
Section~\ref{sec_del}.
In Section~\ref{sec_one} we gather in information on each of the
known \dss\ and in Section~\ref{sec_dis} we give a general discussion
of our results before giving our conclusions in Section~\ref{sec_con}.

\section{The observations\label{sec_obs}}

Spectra were obtained with the DFOSC on the Danish 1.54m telescope at
ESO, La Silla during the period May 24th to June 2nd, 1999.
The Echelle mode was used, employing the combination of grism \#9 cross
dispersed by grism \#11. The detector was a Loral/Lesser C1W7 CCD
and was read out via amplifier A at high gain and 
binning by a factor 2 in the $x$-direction perpendicular 
to dispersion of grism \#9 was performed. 
The spectral resolution is $R=4\,300$ and 
the wavelength range is $3\,000-11\,000$~\AA. 


The resolution of the DFOSC is too small for velocity measurements
and abundance studies, but it is a very efficient instrument. The \dss\
are faint ($V\simeq12-13$) and a large 
telescope is needed to carry out high resolution studies.
Flexure problems prevented precise measurements of radial velocities
which could have been used to study membership of the cluster and binarity.

Two out of eight nights were lost due to bad weather. 
The total number of spectra collected for the rest of the observing 
period was 66 for stars in NGC~6134: typically 3-5 spectra for 
each of the six \ds\ and 2-3 spectra of nine selected ``twin'' stars. 
The twin stars were selected from the Str\"omgren photometry, ie.~we
chose stars with similar values of the $b-y$ index and $V$ magnitude.
Except for the \dss\ \#853 and  \#906 (the evolved ones), the
differences between the photometric values of the variable and its twin were
rather small, i.e.\ $\left|\Delta(b-y)\right| < 0.004$ and 
$\left|\Delta(V)\right| < 0.55$ which corresponds
to $\left|\Delta(T_{\rm eff})\right| < 280 $K and $\left|\Delta(\log g)\right| < 0.3$. 
For the two evolved \dss\ the differences were somewhat larger with
$\Delta(T_{\rm eff}) \simeq -430 $K and $\Delta(\log g) \simeq -0.5$, 
i.e.\ both \dss\ have higher values of $T_{\rm eff}$ and $\log g$ 
compared to their twins.

For calibration purposes we also 
obtained a total of 37 spectra of 10 stars in NGC~3680 and 28 spectra of 
bright reference field stars -- the 13 of which were of HD~118646.
The exposure times were 900--1200~s for stars in NGC~6134, 
at least 1200~s for stars in NGC~3680, and 40--300~s for the field stars.

The $\Delta$a-photometry was obtained in a program including
four other open clusters (see Bayer et al.\ 2000). 
The Bochum 61~cm telescope at ESO and the
Helen-Sawyer-Hogg 61~cm telescope at UTSO (Las Campanas) was
used with CCD cameras with a field of view around 4 arc minutes. 
A total of 32 frames were obtained over 6 nights. 
The detailed data analysis will be published elsewhere 
(Paunzen \& Maitzen 2002).

\section{Data reduction\label{sec_red}}

The reduction of the Echelle spectra were complicated by several
factors. The observations took place during bright time.
Hence, scattered Moon light results in a relatively high sky background 
level, especially for the fainter targets. Two examples are 
shown in Figure~\ref{backgr}.

The large flexures in the DFOSC make it necessary to shift the mask
used to extract the 2D spectra from one frame to the next. This also
makes it difficult to combine spectra of the same target to increase
S/N. Often the \vsini\ was measured for each frame and the average
\vsini\ is computed for the individual exposures.
In order to calibrate the wavelength a large number of Th--Ar
spectra were needed, one before and after each exposure. 
We decided to give up attempts to measure radial velocities
at the precision needed to determine membership ($\sigma_V \le 1$ \kms)
and instead to use the telescope time for obtaining spectra of stars.

Due to the long exposure times night sky emission lines appear 
in the spectra and this makes the background determination
more cumbersome. They do contribute in a positive sense by providing
a sort of wavelength calibration.

The stellar image tends to drift on the slit during and between exposures.
The illumination of the entrance slit is therefore changing all the
time, making the instrumental response vary in time.

The extraction of the spectra from the CCD frames was done using
partly IRAF and partly IDL routines developed to overcome 
the insufficiencies of IRAF.

\subsection{Spectral resolution}

The region around the H$\alpha$ line is 
shown in Figure~\ref{spectra} for the six \dss. 
The nominal resolution of the spectrograph using grism \#9 is $R = 4\,300$
with a 1$\buildrel " \over .$0 slit. We were using a slit of 
1$\buildrel " \over .$5 corresponding
to a resolution of $R = 2\,900$, but on some nights the seeing was 
at or below 1$\buildrel " \over .$0, effectively increasing the resolution up
to $R \simeq 5\,000$. This is the maximum $R$ as it corresponds to 
2 pixels on the detector.

\subsection{Flexures of the spectrograph}

\begin{figure} \centering \footnotesize
\includegraphics[width=88mm]{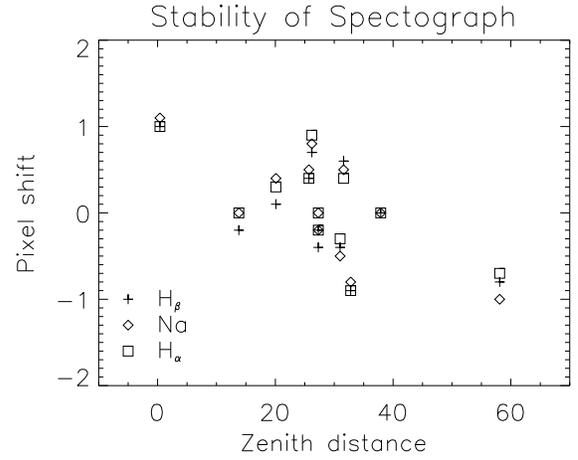} 
\caption{Variations of the position of selected spectral lines for the
    reference star HD~118646 from night to night, sorted by zenith distance. 
    \hbeta\ ($\lambda 4261$\AA) is located in the
    fifth aperture. The Na doublet ($\lambda 5889$\AA) is in aperture \#8. 
    \halpha\ ($\lambda6563$\AA) is in aperture \#9. Estimated uncertainties 
    of the line positions: $\pm 0.1$ pix\label{wav_tab}}
\end{figure}

The effect of the flexures is displayed in Figure~\ref{wav_tab}, which
mainly illustrate our inability to compensate for the flexures in the DFOSC
and the variation in the illumination of the slit. 
Wavelength shifts of the order of 1~\AA\ appear as a result.

\section{Calibration of stellar parameters\label{sec_cal}}

The basic cluster parameters are given in Table~\ref{basic_tab} and are 
based on existing data.
In order to be able to make precise comparisons of the results from
the spectra and the results from the Str\"omgren photometry 
presented by Bruntt et al.\ (1999), a recalibration of the 
photometry has been done.
We have used the \templogg\ code by Rogers (1995) which
automatically selects the most appropriate photometric calibration
based on the Str\"omgren indices.
The program then interpolates the dereddened indices in
the atmospheric \atlasseks\ model grid by Kurucz (1979) and determines the fundamental
atmospheric parameters of the star, ie.~$T_{\rm eff}$, $\logg$, and
[Fe/H]. 
We note that the photometric
calibrations considered by \templogg\ are not the bare results from 
the \atlasseks\ grids, but rather an experimental calibration -- i.e.\ the color grids
are transformed so as to match well-calibrated fundamental stars before the
calibration is made.

We found that the \teff\ values are 
systematically a few hundred degrees lower than what is given in 
Bruntt et al.\ (1999). 
The effect of this change in \teff\ is a change in the age and distance to 
the clusters.
The distances found by Bruntt et al.\ (1999) decrease by 25\% and the 
ages increase by 0.1~Gyr (cf.\ Table~\ref{basic_tab}).

When comparing synthetic spectra to the observations to derive 
rotational velocities we confirmed that the spectra were in
better agreement with the somewhat lower values of $T_{\rm eff}$.

For the \ds\ \#906 (and star \#919, cf.\ Table~\ref{abundance_tab}) the 
$m_1$ index 
is outside the range of the {\tt templogg} program;
the metallicity of this star was derived from the calibration 
of A-type stars by Smalley \cite{smalley}.

A few stars have evolved far off the main sequence and are
outside the range of the \templogg\ code. The \logg\ was then derived by
comparison with stars with the same \teff\ but lying on the main sequence 
by using the
difference in $V$ magnitude and assuming the same bolometric correction.

The parameters that were derived 
for the \dss\ and the known variable blue straggler in NGC~6134 
are listed in Table~\ref{Tab_1}. The average metallicity for NGC~6134 is
\feh\ = $+0.28\pm0.02$ (Bruntt et al.\ 1999); the quoted error 
is the standard deviation of the mean for 50 F-type stars,
and does not include any possible 
systematic errors. We estimate the error on the metallicity 
to be at least 0.10~dex which corresponds to an error on the 
zero point of the $m_1$ index of 10 mmag.
Hence, we will use $[{\rm Fe/H}]_{\rm phot} = +0.28\pm0.10$ from this point.

\begin{table}\centering
  \caption{Basic parameters for the observed open clusters}\label{basic_tab}

  \begin{minipage}{\textwidth} 
   \renewcommand{\thefootnote}{\thempfootnote}

    \begin{tabular}{l|r|r}
      Parameter & \multicolumn{1}{c|}{\ngc{6134}} & \multicolumn{1}{c}{\ngc{3680}}\\
      \hline
      Reddening\footnote{From Bruntt et al.\ (1999)}, $E(b-y)$ & $0.263\pm0.004$   & $0.048\pm0.011$  \\
      Metallicity$^{\text{\thempfootnote}}$, [Fe/H]              & $+0.28\pm0.02$
 & $0.09\pm0.02$  \\
      Cluster distance\footnote{From Bruntt et al.\ (1999); distance decreased by 25\%}, $d$
       & 1.05$\pm$0.06\;\text{kpc} & 0.83$\pm$0.05\;\text{kpc} \\
      Cluster age\footnote{From~Bruntt et al.\ (1999); age decreased by 0.1 Gyr}, $A$                 & 0.79$\pm$0.10\;\text{Gyr} & 1.58$\pm$0.15\;\text{Gyr} \\
      Right ascension, $\alpha$                                  & {$13^{\text h}27^{\text m}42^{\text s}$} &{$11^{\text h}25^{\text m}42^{\text s}$}\\
      Declination, $\delta$                                      & {$-49^\circ 09^{\text m} 00^{\text s}$} & {$-43^\circ 15^{\text m} 00^{\text s}$}\\
    \end{tabular}
  \end{minipage}
\end{table}



\begin{table*}\centering
  \caption{Identification numbers and Str\"omgren indices 
from Bruntt et al.\ (1999) as well as the 
derived astrophysical parameters for the six \dss\ and 
the blue straggler (star \#647) in \ngc{6134}. 
The $c_1$ index was determined from the $c_0$ indices (Bruntt et al.\ 1999). 
The last three columns contain data from the program
\templogg\ (Rogers 1995). The standard errors 
on $T_{\rm eff}$, $\logg$, and [Fe/H] 
are 200~K, 0.2 dex, and 0.2 dex respectively\label{Tab_1}}
  \bigskip
  \begin{tabular}{c|ccccccc|cll}
    \hline\hline
    ID$_3$  &   $V$  & $(b-y)$ & $m_1$  & $c_1$ & $\beta$ & $E(b-y)$ & $(b-y)_0$ & \teff & \logg & \feh \\ 
    \hline
    508 & 13.548 & 0.422   & 0.159  & 0.859 & 2.784   & 0.256    &  0.159    &
7540 &  3.85 & 0.59 \\ 
    574 & 12.452 & 0.461   & 0.104  & 0.902 & 2.738   & 0.271    &  0.198    &
7110 &  3.47 & 0.23 \\ 
    616 & 13.179 & 0.415   & 0.136  & 0.889 & 2.778   & 0.252    &  0.152    &
7470 &  3.79 & 0.36 \\ 
    647 & 11.673 & 0.277   & 0.209  &       & 2.898   &          &  0.014    &
     &       &      \\
    679 & 13.547 & 0.424   & 0.144  & 0.806 & 2.758   & 0.236    &  0.161    &
7320 &  3.86 & 0.46 \\ 
    853 & 11.973 & 0.477   & 0.120  & 0.967 & 2.742   & 0.264    &  0.214    &
6660 &  3.32 & 0.49 \\ 
    906 & 12.266 & 0.486   & 0.181  & 0.802 & 2.698   & 0.227    &  0.223    &
6780 &  3.47 & 1.01 \\ 
    \hline \hline
  \end{tabular}
\end{table*}



\section{Rotational velocities\label{sec_rot}}

The rotational velocities were derived by comparing the observed 
spectrum with a synthetic spectrum where temperature, \logg, and metallicity 
was determined from the Str\"omgren indices. The synthetic spectrum 
$F_{\lambda,synth}$ was convolved 
with the instrumental response $P_{\lambda, instr}$ and the 
rotational broadening function $G_{\lambda}(\vsini)$, thus:
\begin{equation}
F_{\lambda,obs} = F_{\lambda,synth} \otimes P_{\lambda, instr} \otimes G_{\lambda}(\vsini)
\end{equation}

We then determine \vsini\ by minimizing the residuals when subtracting 
the observed and convolved synthetic spectrum. 
In the following we will discuss in detail the technique 
we have used and the results.

\subsection{The synthetic spectra}

The synthetic spectra were supplied by Werner W.\ Weiss, Vienna, calculated
using a modified version of \atlasni\ (Kupka 1996) and the 
line lists from the \vald\ database (Kupka et al.\ 1999). 
The grid was a coarse grid with the following properties:

\begin{itemize}
\item $T_{\rm eff} \in [5\,800, 8\,000]$ K with $\Delta T = 250$~K.
\item $\logg \in [3.80, 4.50]$ with $\Delta \logg = 0.5$
\item $\feh\ \in [0.00,0.50]$ with $\Delta \feh = 0.5$
\end{itemize}
The wavelength range is 4000--8000~\AA\ and the resolution is 0.1~\AA.
For a given star a spectrum was interpolated in this grid to the 
calibrated parameters using the following weights:
\begin{equation*}\label{eq:weights}
  W_i=\left(1-\frac{\Delta(\teff)_i}{250}\right) \!
  \left(1-\frac{\Delta(\logg)_i}{0.5}\right) \!
  \left(1-\frac{\Delta([{\rm Fe/H}])_i}{0.5}\right)
\end{equation*}

\subsection{The instrumental profile}

The seeing was generally good and smaller than the slit width. 
The mean instrumental
profile was estimated by a Gaussian profile. The width of the Gaussian 
was determined by 
calculating \vsini\ for a slowly rotating star HD~118646 with a well
determined $\vsini = 35.3 \pm 1.7$ km~s$^{-1}$ (mean of the
results from Glebocki \& Stawikowski 2000). 
The seeing for a given spectrum was estimated by measuring 
the width of the spectrum perpendicular to the dispersion
direction. This width determines the width of the Gaussian 
instrumental profile 
and gives consistent results for the reference star HD~118646.
Further tests were possible using a few other reference stars 
with known \vsini.

The stars in NGC~3680 are slow rotators due to the relatively 
higher age of the cluster (compared to NGC~6134) and the 
spectral type of the observed stars (F-type stars), 
ie.\ magnetic braking has had time to reduce the rotation. 
Unfortunately the observed stars around the turn-off region
are quite faint and the spectra have poor S/N.
Due to the wide instrumental profile, \vsini\ values below 25~\kms\ are not 
considered: the value is just quoted as being below 25~\kms.

\subsection{Rotational broadening}

We follow Gray (1992) and use for $G_{\lambda}(\vsini)$:
\begin{equation}\label{eq:grayrotation}
G = 
  \frac{2(1-\epsilon)[1-(\Delta\lambda/\Delta\lambda_{\text L})^2]^{\frac{1}{2}} +
    \frac{1}{2}\pi\epsilon[1-(\Delta\lambda/\Delta\lambda_{\text L})^2]}
  {\pi\Delta\lambda_{\text L}(1-\epsilon/3)},
\end{equation}
where
\begin{equation}
\Delta\lambda = v = (\lambda/c) \, x \Omega \, \sin i,
  \label{eq:dopplershift}
\end{equation}
and 
\begin{equation}
 \Delta\lambda_{\text L} = (\lambda/c) \, v \, \sin i.
\end{equation}
A limb-darkening law is used in the form:
\begin{equation}
  I_c/I^0_c = 1 - \epsilon + \epsilon \cdot \cos \theta,
  \label{eq:gray_limb}
\end{equation}
where $\theta$ is the angle formed by the line of sight and the
direction of the emerging flux. $I^0_c$ represents the intensity
at the stellar disk center. The linear limb-darkening
coefficient is chosen to have a value of $\epsilon=0.5$.
Using a constant value has a negligible effect on the results.

\subsection{Spectral lines used for determination of \vsini}

The rotational velocity is determined from a small set of individual
spectral lines.
For the slowly rotating stars we have the choice of using a set of
narrow metal lines and the strong Hydrogen line profiles, which
are wide but have deep core absorption. 
For the faster rotators the metal lines become very shallow and
blended. The influence of the continuum fit begins to dominate
the accuracy and only the Hydrogen lines give good results.

In Table~\ref{line_tab} we give the list of the metal lines that were 
used to measure \vsini\ in HD~118646.

The strong Na doublet is perturbed by lines from intervening
interstellar lines and cannot be used. 

\subsection{Rotation of stars in NGC 6134}


The rotational velocities for the \dss\ and their twins 
in NGC~6134 are given in Table~\ref{Tab_rot}. 
Note that a twin star can be a twin for more than one \ds\ and
some twins appear more than once in Table~\ref{Tab_rot}. 
A \ds\ can also have more than
one twin. There are seven twins for six $\delta$ Scuti stars.
Star \#647 is not a \ds\ but a variable 
blue straggler, and it is included here for completeness. 

It seems that that there is no evidence for any difference
in terms of rotational velocity between \dss\ and stable stars.
This is not a strong statement due to the small number of stars
involved, but it is still interesting.

Our result that the \vsini\ of the \dss\ are evenly distributed
from 30 to 180 \kms\ is in good agreement with Solano \& Fernley (1997).
We find the same distribution for the non-variable stars but here Solano
\& Fernley find a larger spread in rotational velocity. The most
likely reason for this is that we have a small sample of stars 
compared to Solano \& Fernley. 

\begin{table}\centering\footnotesize
    \caption{A list of 21 metallic lines in HD~118646 with sufficient
    width and isolation to be usable for determination of \vsini.
    The table is divided by horizontal lines which correspond to 
    the location of the three orders of the spectrograph that we have used.
    The line position in the spectrum has not been corrected for heliocentric
    motion. The equivalent widths (\ew{}) depend heavily on the
    determination of the continuum\label{line_tab}}

  \begin{tabular}{cclc}
  \hline\hline
  $\lambda^{\text{synth}}$/\AA  &   $\lambda^{\text{obs}}$/\AA    &   Line(s)   &    \ew{} / m\AA \\ \hline
  4957.489 & 4958.29 & \ion{Fe}{i}/\ion{Fe}{i} & 293 \\
  5183.608 & 5184.82 & \ion{Mg}{i} ($\lambda 5183.604$) & 495 \\
  5226.913 & 5227.99 & \ion{Ti}{ii}/\ion{Fe}{i}/\ion{Fe}{ii}/\ion{Fe}{i} & 512 \\
  5328.290 & 5329.59 & \ion{Fe}{i}/\ion{Cr}{i}/\ion{Fe}{i} & 434 \\
  5446.915 & 5447.51 & \ion{Fe}{i} ($\lambda 5446.916)$ & 275 \\
   & 5476.65 & \ion{Fe}{i}/\ion{Fe}{i}/\ion{Ni}{i} & 354 \\ \hline
   & 5447.51 & \ion{Fe}{i} ($\lambda 5446.916)$ & 275 \\
   & 5477.87 & \ion{Fe}{i}/\ion{Fe}{i}/\ion{Ni}{i} & 288 \\
  5528.416 & 5529.18 & \ion{Sc}{ii}/\ion{Mg}{i} & 284 \\
  5615.564 & 5616.69 & \ion{Fe}{i}/\ion{Fe}{i} & 236 \\
  5763.676 & 5754.90 & \ion{Fe}{i}/\ion{Fe}{i}/\ion{Si}{i} & 210 \\
  5762.953 & 5764.10 & \ion{Fe}{i}/\ion{Si}{i} & 190 \\   \hline
  6102.764 & 6103.95 & \ion{Fe}{i}/\ion{Ca}{i}/\ion{Fe}{i}/\ion{Fe}{ii} & 222 \\
  6122.235 & 6123.48 & \ion{Ca}{i} ($\lambda 6122.217$) & 159 \\
  6136.750 & 6138.35 & \ion{Fe}{i}/\ion{Fe}{i}/\ion{Fe}{i} & 254 \\
   & 6142.96 & \ion{Ba}{ii}/\ion{Si}{i} & 140 \\
   & 6163.49 & \ion{Ca}{i}/\ion{Ca}{i} & 273 \\
  6169.349 & 6170.93 & \ion{Ca}{i}/\ion{Ca}{i}/\ion{Fe}{i} & 200 \\   \hline
   & 6279.44 & \ion{Sc}{ii}/\ion{Fe}{i} & 449 \\
   & 6457.38 & \ion{Ca}{i}/\ion{Fe}{ii}/\ion{Si}{i} & 204 \\
   & 6496.52 & \ion{Ca}{i}/\ion{Fe}{i}/\ion{Fe}{i}/ & \\
   &   --      & \quad \ion{Fe}{i}/\ion{Ba}{ii}/\ion{Ca}{i} & 646 \\
  \hline\hline
  \end{tabular}
\end{table}

\begin{table}\centering \footnotesize

\caption{Mean values of \vsini\ (in km~s$^{-1}$) for the variable stars
and their twins in this study. Star \#647 is a blue straggler.
The errors quoted are the standard deviation of the mean\label{Tab_rot}}
\bigskip

\begin{tabular}{cpcp}

\hline\hline

\ds\ & \multicolumn{-1}{c}{\vsini} & Twins &  \multicolumn{-1}{c}{\vsini} \\

\hline
508  &    33+8   & 218        &    35+7    \\
574  &    60+5   & 754        &    90+16   \\
     &             & 919        &    88+6    \\
     &             & 969        &   157+15   \\ 
616  &   163+14  & 689        &    90+11   \\
     &             & 865        &   115+14   \\
647  &   120+25  &            &            \\
679  &   145+13  & 218        &    35+7    \\
853  &   112+9   & 969        &   157+15   \\
906  &    50+7   & 830        &   105+18   \\
\hline\hline

\end{tabular}

\end{table}


\begin{figure*}\centering
  \includegraphics[width=\textwidth,height=7cm]{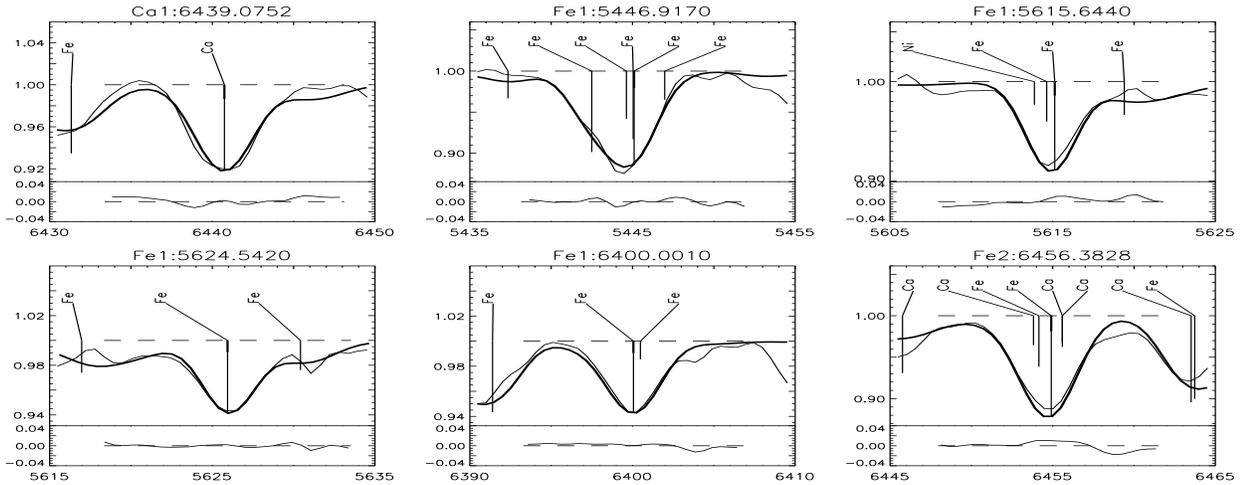}
  \caption[fitted_lines_n6s574]{The observed (thin line) and fitted synthetic 
spectrum (thick line) for six lines for the \ds\ \#574. The lower portion
of each plot shows the relative difference of the observed and 
synthetic spectrum. The position of the 
fitted line and the deepest neighbouring lines are marked by
vertical lines. The length of these lines are scaled with the
theoretical line depth of the fitted line\label{fittedlines}}
\end{figure*}

\begin{figure} \centering \footnotesize
 \includegraphics[width=85mm]{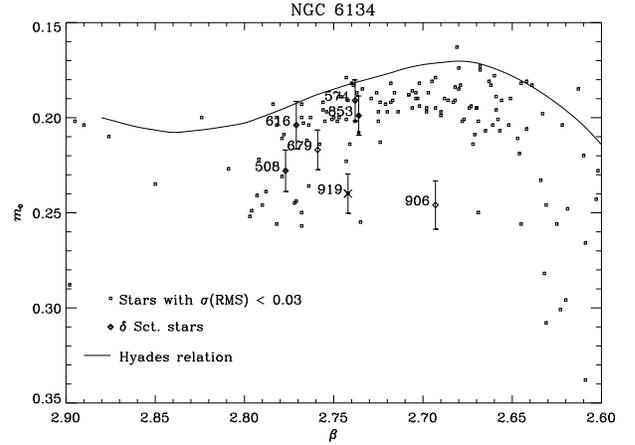}
 \caption{The relation between $m_0$ index and $\beta$ for
 stars in NGC~6134. The solid line is the Hyades relation
 adapted from Crawford (1975a, 1979). The error bars on $m_0$ 
 are shown for the $\delta$ Scuti stars and the 
 metal rich non-variable star \#919; the plotted errors 
 include a systematic error of 0.02 mag. 
 Note that only stars with photometric
 errors on $m_0$ and $\beta$ below 0.03~mag are shown \label{m1-index}}
\end{figure}


\subsection{Rotation of other stars}

Among the 10 stars we observed in NGC~3680 most of them
have $V \simeq 14-15$ and were thus fainter than our targets 
in NGC~6134 where the range among the 15 observed 
stars is $V \simeq 12-13$. 
Thus the S/N was much lower for the
stars in NGC~3680 and the determination of the $v \sin i$ is more uncertain.
In summary we find that our sample of stars in NGC~3680 are 
all slow rotators and only an upper limit of 25 \kms\ could 
be determined -- a limit
which is due to the resolution of the spectrograph.

We also observed several reference stars. Most of them
have $\vsini$ below 25 \kms\ and the exceptions are:
for HD~118646 we find $\vsini = 28.5\pm1.6$, HD~132475 has
$\vsini = 35\pm10$, and for HD~155458 we find $\vsini = 105\pm20$.
For HD~118646 the mean of three measurements listed in
Glebocki \& Stawikowski (2000) is 
$v\sin i = 35.3\pm1.7$ \kms\ which is in rough agreement with what we find. 
The value of $v\sin i = 11\pm6$ \kms\ for HD~132475 quoted
in Glebocki \& Stawikowski (2000) is from Stetson (1983).
This is calculated using a method which depends on the calibration
of a single star (HD~24450) which has a poorly known value
of \vsini\ (see p.\ 1362 in Stetson 1983) and this may be 
the cause for the disagreement with our result 
result of $v\sin i=35\pm10$ \kms.

The method was also tested on two spectra obtained with different
instruments. A spectrum from the ELODIE spectrograph 
at $R \approx 48\,000$ of the \ds\ HD~49434
was analysed. We found $\vsini = 70$~\kms\ using \halfa\ and
$\vsini = 80$~\kms\ using ${\rm H}_{\beta}$.
This is in good agreement with the value of $\vsini = 84$~\kms\ measured by
Bruntt et al.\ \cite{hd49434}.

To summarize, our results for the field stars 
HD~118646 and HD~49434 agree with independent determinations, and confirm
the validity of the \vsini\ produced by our technique.


\begin{table*}\centering
  \caption{
Derived abundances for four \dss\ stars 
and the twin star \#919 in NGC6134,
reference star HD~118646, and the Sun (from Grevesse et al.\ 1996). 
The first column is the identification
number used by Bruntt et al.\ \cite{bruntt2}. 
The numbers in parenthesis
give the internal RMS error and $n$ is the number of lines used. The
microturbulence that was used for the computation of the synthetic
lines is given -- the approximate error is 0.6 \kms. The last columns
give the metallicity determined by the $m_1$ index and 
spectroscopy\label{abundance_tab}}
  \bigskip

\begin{minipage}{\textwidth} 
\renewcommand{\thefootnote}{\thempfootnote}
\centering
  \begin{tabular}{l|lc|lc|c|cl|c}

    ID$_3$   &  $\log N_{\rm Fe}/N_{\rm tot}$  & $n$   &    $\log N_{\rm Ca}/N_{\rm tot}$  &  $n$  & $v_{\rm micro}$ [\kms] & $m_1$ & $[{\rm Fe/H}]_{\rm phot}$ & $[{\rm Fe/H}]_{\rm spec}$  \\
\hline
508\footnote{Possibly an Am star due to high [Fe/H] and low [Ca/Fe], see Conti 1970}
         & -4.05(0.21)& 16    & -5.38(0.24)&   6   & 2.3 & 0.190 & 0.59(0.11) & 0.49(0.15) \\ 
574      & -4.30(0.17)& 12    & -4.99(0.49)&   7   & 2.5 & 0.104 & 0.23(0.11) & 0.24(0.15) \\ 
679      & -4.16(0.27)&  7    & -4.68(0.39)&   2   & 2.3 & 0.144 & 0.46(0.11) & 0.38(0.17) \\ 
906      & -4.12(0.34)& 10    & -4.65(0.09)&   4   & 1.4 & 0.181 & 1.01(0.15)
 & 0.42(0.18) \\ 
919\footnotemark[\value{mpfootnote}]
         & -4.09(0.28)& 11    & -5.54(0.46)&   5   & 2.0 & 0.173 & 0.90(0.12) & 0.45(0.16) \\ 
\hline
HD118646 & -4.44(0.20)& 12    & -5.32(0.17)&   9   & 1.2 & --     & 0.00(0.09) & 0.10(0.15) \\ 

\hline
Sun      & -4.54      & --     & -5.68      &   --   & --   & --     & 0.00       & --          \\ 

  \end{tabular}

 \end{minipage}

\end{table*}


\section{Abundances\label{sec_abu}}

The average metallicity of NGC~6134 was determined from
the $m_1$ index of the F-type stars by Bruntt et al.\ (1999), 
ie.\ [Fe/H]$_{\rm phot}=+0.28\pm0.10$. 
In addition Bruntt et al.\ (1999) found that several of the A-type 
stars have a very high $m_1$ index, indicative of a high metallicity.
The reddening corrected $m_0$ index is shown 
versus the temperature sensitive $H_\beta$ index in Figure~\ref{m1-index}
The turn-off stars lie around $\beta = 2.77\pm0.02$ and temperature 
increases to the left.

For two of the \dss\ a quite high $m_1$ index has been measured
(star number \#508 and \#906, cf.\ Table~\ref{Tab_1}).
The high abundances of metals in A-type stars is
thought to be an effect of diffusion in the upper layers
present in slowly rotating stars with stable envelopes, which
would tend to decrease also the content of Helium in the
driving layers and make the star stable. 
Interestingly, high metallicity and
variability is suspected to be exclusive phenomena (Kurtz 2000).

In order to confirm this we 
decided to test the \feh\ results from the 
Str\"omgren photometry by doing abundance analysis
by using the low resolution spectra obtained to measure \vsini.
We calculated the atmosphere models with 
a modified version of the \atlasni\ code (Kupka 1996),
and extracted line lists from \vald\ (Kupka et al.\ 1999). 
We have developed software called \vwa\ for semi-automatic 
abundance analysis which is described in Bruntt et al.\ \cite{hd49434}:
\vwa\ automatically selects lines that suffer from the 
the least amount of blending. For a region around each 
selected line the synthetic spectrum is calculated for 
different abundances of the element forming the line 
until the equivalent width of the observed and synthetic 
spectrum match. 
When this is done, each fitted line is inspected visually -- and
the fit may either be accepted, rejected or improved manually.

The abundance analysis did not produce reliable results for
all the stars. For most of the non-variable stars only two 
spectra were obtained, hence the S/N is small 
and the errors on the derived abundances are large.
Since we obtained more 
spectra of the \dss\ in NGC~6134 we were able to
make the best abundance determination for some of these stars.
The exceptions are the \dss\ \#616, \#647, and the 
blue straggler \#853 for which the combination of 
high \vsini\ and low S/N made the abundance analysis 
impossible.

Examples of the result of the automatic fit of abundances 
are shown in Figure~\ref{fittedlines} for a Ca-line and five Fe-lines 
as seen in the spectrum of the \ds\ \#574. This star has
quite low $\vsini$, hence it is possible to find a number
of Fe-lines that are only partially blended.


The abundances determined with \vwa\ from the Fe and Ca lines, 
are presented in Table~\ref{abundance_tab}.
The abundance of element $X$ is given as $\log N_{X}/N_{\rm tot}$ where
$N_X$ is the number of atoms of element $X$ and 
$N_{\rm tot}$ is the total number of atoms in the atmosphere model.
The error quoted in parenthesis in Table~\ref{abundance_tab} is the
RMS internal error. 

To get a realistic idea of the error on the
abundances one must consider the uncertainty of the 
parameters used for the computation of the model atmosphere and
the synthetic spectrum, 
most noticeably $T_{\rm eff}$ and microturbulence. We assume that 
a realistic error estimate of $T_{\rm eff}$ is 200~K 
which corresponds to an error of about 0.1 dex in the
Fe abundance.
For the \dss\ we have used microturbulence in the range 1.4--2.5 \kms. 
We estimate the uncertainty to be around 0.6 \kms\ which corresponds 
to an error in the Fe abundance of about 0.1~dex. 
For the final result we have combined the 
uncertainties due to the internal scatter (sd.\ of the mean)
and the contributions from the estimated error 
on $T_{\rm eff}$ and microturbulence.

The derived metallicities from photometry and spectroscopy 
are given in the last two columns of Table~\ref{abundance_tab}.
The quoted errors include a 10 mmag zero point error in the $m_1$ index 
(for $[{\rm Fe/H}]_{\rm phot}$) and uncertain model 
parameters (for $[{\rm Fe/H}]_{\rm spec}$) as was discussed above.

Abundance analysis was possible for four \dss\ in NGC~6134 
and the mean Fe abundance is
$\log N_{\rm Fe}/N_{\rm tot} = -4.16 \pm 0.05$~(the quoted error 
is the sd.\ of the mean).
Assuming that these
stars have the same H and He abundances as the Sun we estimate 
the average metallicity of A-type stars in NGC~6134 to be
$[{\rm Fe/H}]_{\rm spec} = +0.38\pm0.05$. This is in good agreement with
the result from Str\"omgren photometry (F-type stars) which 
is $[{\rm Fe/H}]_{\rm phot} = +0.28\pm0.10$. 

The \dss\ \#508, \#906, and the twin star \#919 are 
found to be metal rich from the Str\"omgren 
photometry and this is confirmed by the spectroscopic results.
For the stars \#508 and \#919 we find a low
[Ca/Fe] ratio which is characteristic for Am-type stars.
We must stress that only few usable Ca lines could be found 
for abundance analysis, and hence the abundance is quite uncertain. 
Still, the result is interesting since our results 
indicate that the \ds\ \#508 is then
possibly an Am-type star and at the same time a \ds.

\section{$\Delta$a-photometry\label{sec_del}}

In this section we present results for $\Delta$a-photometry in order
to shed more light on the true nature of the bona-fide Am-type candidates
and the variable objects within NGC~6134. The $\Delta$a-system is 
capable to detect chemically peculiar stars (most efficient for magnetic
CP2 stars in the notation of Preston 1974) via three narrow-band filter
photometry. It samples the broad band flux depression around 520\,nm
in three filters: $g_{1}$ at 500\,nm and $y$ at 550\,nm, both representing 
the continuum at the edges of the feature, and $g_{2}$ located at the center 
of the depression at 521\,nm. The $\Delta$a magnitude is the difference of
$g_{2}$ relative to the mean of $g_{1}$ and $y$, normalized to zero for 
a non-peculiar object: $a=g_{2} - (g_{1}+y)/2$. The photometric
accuracy needed to detect classical peculiar stars is typically a few mmags
(Maitzen 1976).
An $a$ versus ($g_{1}-y$) diagram
can therefore be used to detect peculiar objects and to sort out fore- and
background stars due to a high sensitivity to interstellar reddening.
The standard relations are
\begin{eqnarray}
E(g_{1}-y) & = & 0.54 \cdot E(b-y) \label{eq1} \\
f & = & {E(a)}/{E(b-y)} \label{eq2}
\end{eqnarray}
with $f$ being of the order of 0.05. If an open cluster does not suffer
from differential reddening (such as NGC~6134), the normality line
has a very small intrinsic scatter. Objects with a significant
different reddening than the mean of the cluster, will be shifted almost 
horizontally in the $a$ versus ($g_{1}-y$) diagram (see for
example the case of IC~2602 in Maitzen 1993).


\begin{figure} 
\begin{center}
\includegraphics[width=66mm,height=100mm]{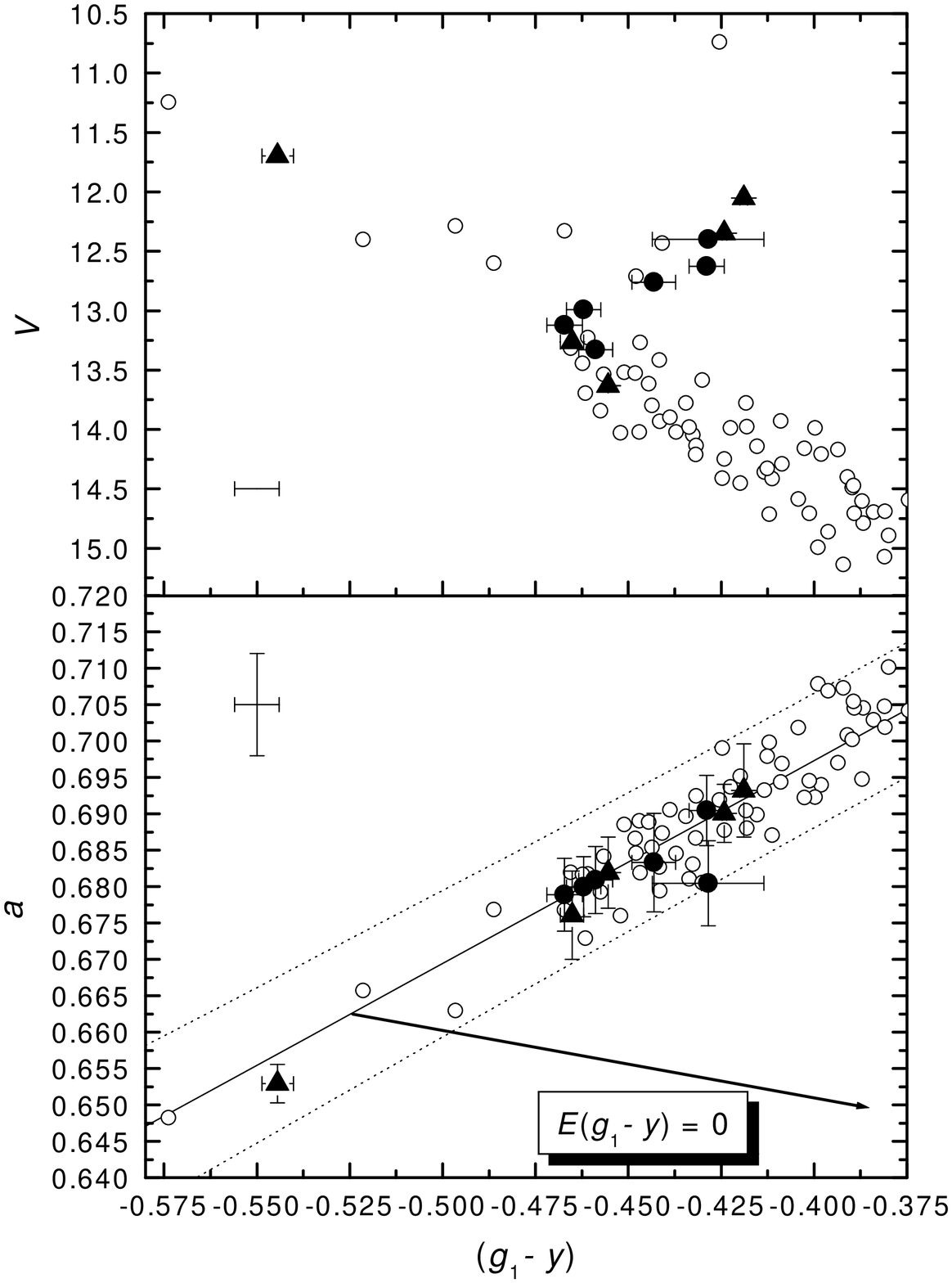}

\caption{$V$ and $a$ versus ($g_{1}-y$) diagrams for NGC~6134 with 
the reddening vector for the temperature range from A0 to F5. 
Filled circles indicate
bona-fide Am-type candidates whereas filled triangles are known variable stars
from the literature. The full line is the
normality line whereas the dotted lines are the confidence intervals
corresponding to 99.9 percent. The error bars are shown for selected
individual stars and the mean error bars are shown to the left}  
\label{fig_g1y}
\end{center}
\end{figure}


What is the behaviour of Am-type stars in the $\Delta$a photometric system?
Maitzen (1976) and Vogt et al.\ (1998) investigated a sample of these
objects. They found that the mean $\Delta$a-values are 6\,mmag above the
normality line with some very peculiar Am-type stars reaching 14\,mmag.
On the other hand classical CP stars have $\Delta$a of up to 80\,mmag.

The measurements for NGC 6134 are presented in Table~\ref{res_g1y}.
The errors on the variable stars are unfortunately too large to
measure a $\Delta$a $\ge 0.006$~mmag. 

\begin{table*}[t]
\caption{The results of the $\Delta$a-photometry for 
bona-fide Am-type candidates (upper panel) and known 
variable stars (lower panel)}
\label{res_g1y}
\begin{center}
\begin{tabular}{crrrrccccc}
\hline
\multicolumn{1}{c}{$ID_{1}$} & \multicolumn{1}{c}{$ID_{2}$} & 
\multicolumn{1}{c}{$ID_{3}$} &
\multicolumn{1}{c}{$V$}&\multicolumn{1}{c}{$a$}&
\multicolumn{1}{c}{$(g_{1}-y)$}&
\multicolumn{1}{c}{$\Delta a$}&
\multicolumn{1}{c}{$(b-y)$}&\multicolumn{1}{c}{$\Delta$a} &
\multicolumn{1}{c}{$N$}\\   
\hline
5	& 106 &	832	& 12.400	& 0.680(6)	& $-$0.428(15)	& $-$0.009	& 0.452	& $-$0.008	& 5  \\
23	& 50 &	1023	& 12.762	& 0.683(7)	& $-$0.443(6)	& $-$0.002	& 0.434	& $-$0.002	& 30  \\
27  & 61 &	598	& 12.992	& 0.680(4)	& $-$0.462(5)	& +0.000	& 0.423	& $-$0.004	& 32 \\
48	& 9	& 689	& 13.122	& 0.679(5)	& $-$0.467(5)	& +0.000	& 0.417	& $-$0.004	& 32 \\
62	& 4	& 619	& 13.327	& 0.681(5)	& $-$0.459(5)	& +0.000	& 0.422	& $-$0.003	& 32 \\
75	& 41 &	919	& 12.628	& 0.690(5)	& $-$0.429(5)	& +0.001	& 0.460	& +0.001	& 30 \\
\hline
9	& 54 &	906	& 12.347	& 0.690(4)	& $-$0.424(4)	& +0.000	& 0.486	& $-$0.003	& 30  \\
10	& 109 & 647	& 11.697	& 0.653(3)	& $-$0.544(3)	& $-$0.004	& 0.262	& $-$0.006	& 30  \\
22	& 59 &	679	& 13.631	& 0.682(5)	& $-$0.455(5)	& +0.000	& 0.424	& $-$0.002	& 32  \\
38	& 66 &	&	13.264	& 0.676(6)	& $-$0.465(6)	& $-$0.003	&	&	& 32  \\
57	& 15 &	853	& 12.051	& 0.693(6)	& $-$0.419(6)	& +0.001	& 0.477	& +0.001	& 30  \\
85	& 21 &	735	& 14.452	& 0.695(5)	& $-$0.420(4)	& +0.003	& 0.483	& +0.002	& 28  \\
\hline
\\
\multicolumn{10}{l}{Col. 1: Identification from Paunzen \& Maitzen(2002)} \\
\multicolumn{10}{l}{Col. 2: Identification from Lindoff (1972)} \\
\multicolumn{10}{l}{Col. 3: Identification from Bruntt et al.\ (1999)} \\
\multicolumn{10}{l}{Col. 4: Visual magnitude} \\
\multicolumn{10}{l}{Col. 5, 6: mean $a$-index and its standard deviation} \\
\multicolumn{10}{l}{Col. 7, 8: mean $(g_{1}-y)$ value and its standard deviation} \\
\multicolumn{10}{l}{Col. 9: Deviation from cluster line ($\Delta$a)} \\
\multicolumn{10}{l}{Col. 10: $(b-y)$ from the literature} \\
\multicolumn{10}{l}{Col. 11: Deviation from cluster line ($\Delta$a)} \\
\multicolumn{10}{l}{Col. 12: Number of measurements} \\
\end{tabular}
\end{center}
\end{table*}

The normality lines based on both ($b-y$) and ($g_{1}-y$) have
been determined from the observed sample of stars:

$$ a_0  =  0.618(3) + 0.155(5) \cdot (b-y)  $$
and 
$$ a_0  =  0.809(2) + 0.279(6) \cdot (g_1-y). $$ \\

The zero point of these relations are shifted less
than 1 mmag compared to relations derived for 
less metal rich clusters. Likewise, the slopes we have found
here or similar to what we have found for the other clusters we
have studied (Paunzen \& Maitzen 2002).
The 3$\sigma$ detection limit was determined to be 9 mmag. 

This limit is quite high and will not allow us to detect weak
Am-type stars. The reason for the ``high'' detection limit at 9 mmag
is that the normality line is not known with sufficient accuracy 
due to the combination of the photometric accuracy, 
the spread in metallicity and intrinsic chemical peculiarity, 
and possible non-uniform reddening across the field of NGC~6134.

From the correlation of 84 stars in common with 
Bruntt et al.\ (1999) we find the transformation of 
our $y$ measurements to standard $V$ values as:
$$ V = -8.138(12) + y. $$

Figure~\ref{fig_g1y} shows the $V$ and $a$ versus ($g_{1}-y$) diagrams.
The filled circles are the measured bona-fide Am-type candidates, ie.\ star 
number 598, 619, 832, 919, and 1023, whereas the filled triangles are the 
known variable stars from the literature: star number 66 (number 
according to Lindoff 1972), 647, 679, 853, and 906. 
The lower panel of Figure~\ref{fig_g1y}
includes the reddening vector corresponding to the mean of the 
cluster $E(b-y) = 0.263$ mag
and the relations \ref{eq1} and \ref{eq2}. Table~\ref{res_g1y} summarizes
the results of all relevant objects. From Figure~\ref{fig_g1y} we may
conclude the following about the investigated objects (cf.\ Table~\ref{res_g1y}): 
\begin{itemize}
\item they are true members of NGC~6134
\item they do not show any significant enhancement of the 520\,nm depression
typical mainly for classical magnetic CP stars.
\end{itemize}
We are not able to definitely rule out an Am-type nature but it seems rather 
unlikely since all objects exhibit almost zero or even slightly negative
$\Delta$a-values.

Five additional variables have been identified during the
$\Delta$a-photometry with rather high level of variability compared
to the \dss\ discussed here. The identification numbers are
3 (number from Lindoff 1972), 735, 1046, 1115, and 
1154 (the last four star numbers according to Bruntt et al.\ 1999). 
Due to the distribution of the observations, 
no conclusion about the nature of variability
can be drawn. In the relevant temperature range one might
think of $\delta$ Scuti and $\gamma$ Doradus pulsation as well as
variability due to eclipsing binaries. 
Further photometric
observations are needed to clarify the true nature of the variability
of these five stars.

\section{The variables one by one\label{sec_one}}

Below we will describe the stars one by one.  The detailed
information on the known oscillation modes can be found
in Viskum et al.\ \cite{ngc6134_dss_frequencies}.

\subsection{Star \#508}

This star is one of two faint \dss\ close to the main sequence.
It is thus a bit surprising to find this star rotating slower
than all other \dss\ in the cluster, but it might of course be
due to the inclination $i$ being close to zero. Two modes were detected
with low amplitudes at $\sim1.5$~mmag. The result from Str\"omgren photometry
and spectroscopy say that the star has a
high \feh\ (cf.\ Table~\ref{abundance_tab}). 
The $v\sin i = 33\pm 8$ is derived from
six spectra. Due to the high metallicity \hbeta\ was too
contaminated to be used and the rotation is measured using
only the \halpha\ line. The star could very well be a marginal
Am star. Its \feh\ value is increased, but its [Ca/H] value is
lower than the mean for the other stars. This is the characteristics
for Am-type stars (see Wolff 1983, chapter 5). It might be a star 
similar to HD~1097: an Am-type star which is pulsating with low 
amplitudes
and was discovered by Kurtz (1989, 2000). Unfortunately $\Delta$a
photometry was not made for this star.

\subsection{Star \#574}

This star is slightly evolved. Only a short time string was
observed and only two modes could be detected with the 
coarse frequency resolution. The star rotates with a modest
\vsini\ = $60\pm 5$. Four spectra were analysed and
both \halpha\ and \hbeta\ lines were used. The metallicity 
is similar to the cluster mean as determined from 
spectroscopy and photometry.

\subsection{Star \#616}

Both \halpha\ and \hbeta\ could be used in this star, but only
one spectrum was observed. The lines are broad leading to 
the highest \vsini\ = $163\pm 14$ among 
the $\delta$ Scuti stars in NGC 6134. The star
is only slightly evolved and is located exactly at the turnoff point
of the cluster. Three modes have been identified, but
more modes are probably present. The $m_1$ index is similar
to the cluster mean, but an abundance analysis could not be made
because of the high \vsini\ and the low signal to noise.

\subsection{Star \#679}

This is the second main sequence star with three modes identified
at low amplitudes at 1--2~mmag. Five spectra were analysed
and they indicate that the star is a fast rotator 
$\vsini = 145\pm 13$ using the \halpha\ line.
The S/N was good when combining the five spectra, so we could
make an abundance analysis, but the \vsini\ is very high for this star
so it is difficult to find good lines. 
Still, the derived metallicity agrees 
with the $m_1$ index, ie.\ the star has slightly higher metallicity
compared to the cluster mean.

\subsection{Star \#853}

This is the most evolved \ds\ with only a single relatively 
high amplitude mode identified with $A = 6.2$~mmag. It is
rotating at $v \sin i = 112\pm 9$ which is 
derived from both the \halpha\ and \hbeta\ lines.
The $m_1$ index is higher than the mean for the cluster, 
but we could not confirm this from the spectra of the star, 
since the S/N was too low.

\subsection{Star \#906}

This is also an evolved star with five low frequency modes in 
the range 60--90~$\mu$Hz, where four modes are very closely spaced
within 6~$\mu$Hz. The interpretation of the
close frequencies as rotationally split modes is supported
by the small $v \sin i = 50\pm7$, mainly derived from
\halpha. A high value of the $m_1$ index suggest that the
star has a high metallicity, and the spectroscopic value
for \feh\ is in agreement, but showing a more modest value.
It does not seem to be an Am-type star as the $\Delta$a index
and the spectroscopic Ca abundance do not show any deviation
from the cluster mean. 

\subsection{Star \#919}

The star is not a variable, but we include it here since it 
may well be an Am-type star:
the \feh\ is high, but it has a low [Ca/Fe] ratio. The $\Delta$a
observations do not show indications of the star being 
an Am-type star, however.
The projected rotational velocity is $v \sin i = 88\pm6$.

\section{Discussion\label{sec_dis}}

We have found that \dss\ do not seem to be different from their twin stars 
in NGC~6134 in terms of their stellar parameters.
This conclusion is consistent with Solano \& Fernley (1997) who used the
same methods of calculation. Their study did show a difference in the spread
of \vsini\ between low amplitude variables and non-variable stars, 
but the rather
small number of stars in our study may be the cause of this difference.

What do we find in other open clusters? We have collected data similar to
what we have for NGC~6134 from the open cluster database (Mermilliod 1996) and
from the catalogue of rotational velocities for four nearby clusters
(Glebocki \& Stawikowski 2000). Stars for which \vsini\ and Str\"omgren 
indices are available 
were chosen,
and from this list stars with $0.09 < (b-y) < 0.23$ were selected, ie.\ the
A-type stars (the interstellar reddening is negligible 
for the nearby clusters examined here).
The number of stars satisfying these criteria is small. The
number of stars in each of the four open cluster lies 
in the range $14 \le N \le 31$.

\begin{figure}\centering
 \includegraphics[width=88mm]{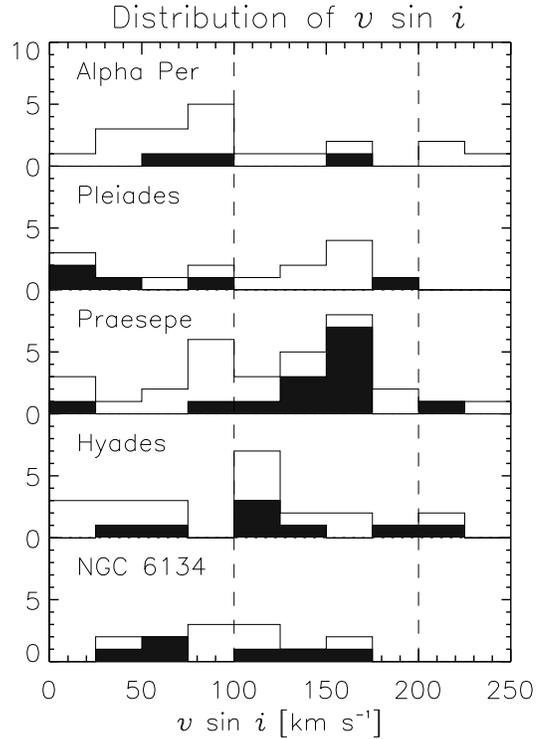} 
  \caption[plot_ha.pro]{Histograms of the distribution of 
rotational velocities in five open clusters. The black columns
is the fraction of \dss\ in a given column, eg.\ in Hyades 3 of 7
stars in the range $100 < \vsini < 125$ \kms are 
$\delta$ Scuti stars\label{vsini}}
\end{figure}

We want to investigate if the distribution of \vsini\ for the \dss\ is 
similar to non-variable stars in several open clusters.
Unfortunately, the question is difficult to answer due to
the small number of stars. Looking at Figure~\ref{vsini} no overall pattern can
be seen. In Praesepe the \dss\ seem to be found almost exclusively among the
fast rotators, but in the Pleiades the slow rotators are variable stars
as well.
In the remaining clusters including NGC~6134 the \dss\ seem not to
be restricted to any particular range of \vsini.

The median of the distributions are presented in Table~\ref{median}.
There is clearly an indication that stars in the nearest clusters 
on average rotate faster than in $\alpha$~Per and NGC~6134.

\begin{table}
\caption{Median of the distribution of \vsini\ for the whole cluster
and the \dss\ for five open clusters}
\centering
\begin{tabular}{l|cc}
            &  \multicolumn{2}{c}{\vsini\, [km\,s$^{-1}$]} \\
\hline
        &    Cluster                       & $\delta$ Scuti \\
Cluster & \multicolumn{1}{c}{stars} & \multicolumn{1}{c}{stars}\\
\hline
$\alpha$~Per & 85 & 75 \\
Pleiades & 115 & 25 \\
Praesepe & 130 & 152 \\
Hyades & 106 & 109 \\
NGC 6134 & 90 & 112 \\
\hline
\end{tabular}
\label{median}
\end{table}

Another aim of the present work is to investigate if there is
any difference in the abundances of \dss\ and non-variable stars.
We find that the \dss\ generally seem to have metallicities similar to 
the cluster mean,
although one of the \dss\ may be a marginal Am-type star. 
Spectroscopy with higher
resolution is needed to confirm these results. If slowly rotating stars
can also be variable, it is not evident that the Am-type
character prevents a star from pulsating. 
Indeed, it is possible that the $\kappa$-mechanism
has not been blocked by diffusion of He from the driving region
in all Am-type or marginal Am-type stars.


\section{Conclusions\label{sec_con}}

We have presented the results of our analysis of spectra 
of variable and non-variable stars in the metal rich open cluster NGC~6134.
We have determined the rotational velocity of several stars
and made abundance analysis of the \dss\ stars.
We summarize our conclusions below:

\begin{itemize}

\item{We have determined \vsini\ for the six known \dss\ and 
a set of known non-variable A-type stars in NGC 6134}

\item{We have determined \feh\ from spectroscopy and find a good 
agreement with the value from the $m_1$ index. 
We confirm that NGC~6134 is quite metal rich 
with \feh\ around $0.3\pm0.1$ (error estimate includes 
systematic errors). 
From the $m_1$ index some stars seem to have enhanced 
atmospheric metal content and could be Am-type stars. 
We find that one of the 
\dss\ seem to be a marginal Am-type star (star \#508)}

\item{We have presented $\Delta$a photometry which was carried out in
order to detect any chemically peculiar stars. We find no 
classical Am-type stars in our sample of stars but we note that we
did not obtain $\Delta$a for the interesting \ds\ \#508}

\item{The distributions of rotational velocities of \dss\ and 
non-variable A-stars in five well studied open clusters
have been analysed: we find that the distributions are quite different
but no safe conclusions can be drawn due to the small number of stars.
For NGC~6134 we find that half of the known \dss\ have low \vsini\ and 
they would be suitable for modeling by the use 
of asteroseismology in the future}

\end{itemize}


\begin{acknowledgements}

We thank Werner W.\ Weiss (Vienna) for providing us with the
grid of synthetic spectra. We are grateful for the Danish 
ESA Science Program ({\em ESA F\o lgeforskning}) grant number 9902474.

\end{acknowledgements}


\end{document}